\documentclass[fleqn,usenatbib]{mnras}

\usepackage[T1]{fontenc}
\usepackage{ae,aecompl}
\usepackage{graphicx}
\usepackage{epstopdf}
\usepackage{amsmath}
\usepackage{amssymb}
\usepackage{enumerate}
\title[Transient Survey Merger Jet Rates]{Transient Survey Rates for Orphan Afterglows from Compact Merger Jets}
\author[G.~P. Lamb, M. Tanaka, and S. Kobayashi]{Gavin P Lamb$^{1,}$$^{3}$, Masaomi Tanaka$^{2}$, Shiho Kobayashi$^{1}$
\\
$^{1}$Astrophysics Research Institute, LJMU, IC2, Liverpool Science Park, 146 Brownlow Hill, Liverpool L3 5RF, 
UK\\
$^{2}$National Astronomical Observatory of Japan (NAOJ), Tokyo, Japan\\
$^{3}$Department of Physics and Astronomy, University of Leicester, University Road, Leicester, LE1 7RH, UK}
\date{Accepted 2018 February 18. Received 2018 February 05; in original form 2017 December 01}
\pubyear{2017}
\begin{document}
\label{firstpage}
\pagerange{\pageref{firstpage}--\pageref{lastpage}}
\maketitle

\begin{abstract}
Orphan afterglows from short $\gamma$-ray bursts (GRB) are potential candidates for electromagnetic (EM) counterpart searches to gravitational wave (GW) detected neutron star or neutron star black hole mergers.
Various jet dynamical and structure models have been proposed that can be tested by the detection of a large sample of GW-EM counterparts.
We make predictions for the expected rate of optical transients from these jet models for future survey telescopes, without a GW or GRB trigger.
A sample of merger jets is generated in the redshift limits $0\leq z\leq 3.0$, and the expected peak $r$-band flux and timescale above the LSST or ZTF detection threshold, $m_r=24.5$ and $20.4$ respectively, is calculated.
General all-sky rates are shown for $m_r\leq26.0$ and $m_r\leq21.0$.
The detected orphan and GRB afterglow rate depends on jet model, typically $16\la R\la 76$ yr$^{-1}$ for the LSST, and $2\la R \la 8$ yr$^{-1}$ for ZTF.
An excess in the rate of orphan afterglows for a survey to a depth of $m_r\leq26$ would indicate that merger jets have a dominant low-Lorentz factor population, or the jets exhibit intrinsic jet structure.
Careful filtering of transients is required to successfully identify orphan afterglows from either short or long GRB progenitors.
\end{abstract}

\section{Introduction}
The most promising candidate for the progenitor of short $\gamma$-ray bursts (GRBs) is the merger of a binary neutron star (NS) system or a NS black hole (BH) system \citep[e.g.][]{1989Natur.340..126E,1992ApJ...395L..83N,1993Natur.361..236M,2007ARep...51..308B,2007PhR...442..166N,2014ARA&A..52...43B}.
Such systems are candidate targets for gravitational wave detectors, and as such there has been a focus on potential electromagnetic (EM) counterparts to such mergers \citep{2013ApJ...767..124N}.
Amongst the counterparts are the isotropic macro/kilo-nova \citep[e.g.][]{1998ApJ...507L..59L,2013ApJ...775...18B,2013ApJ...775..113T,2013MNRAS.430.2121P,2014MNRAS.441.3444M,2014ApJ...780...31T,2017arXiv170809101T,2016ApJ...829..110B,2016AdAst2016E...8T,2017LRR....20....3M}, radio counterparts \citep[e.g.][]{2011Natur.478...82N,2014MNRAS.437L...6K,2015MNRAS.452.3419M,2015MNRAS.450.1430H,2016ApJ...831..190H}, wide angle coccoon emission \citep{2016arXiv161001157L,2017ApJ...834...28N,2017arXiv170510797G,2017arXiv171100243K}, resonant shattering, merger-shock or precursor flares \citep{2013AAS...22134601T,2014MNRAS.437L...6K,2016MNRAS.461.4435M}, GRBs \citep[e.g.][]{2014MNRAS.445.3575C,2016A&A...594A..84G,2017arXiv170807488K,2017arXiv170807008J}, failed GRBs (fGRB) \citep{2000ApJ...537..785D,2002MNRAS.332..735H,2003NewA....8..141N,2003ApJ...591.1097R,2016ApJ...829..112L,2017arXiv170603000L}, and off-axis orphan afterglows \citep[e.g.][]{2002MNRAS.332..945R,2002ApJ...570L..61G,2007A&A...461..115Z,2013ApJ...763L..22Z,2016arXiv161001157L,2017ApJ...835....7S}.
Some of these counterparts make promising potential transients for the next generation of optical survey telescopes e.g. Large Synoptic Survey Telescope (LSST) \citep{2009arXiv0912.0201L}, and Zwicky Transient Factory (ZTF) \citep[e.g.][]{2014htu..conf...27B,2017NatAs...1E..71B}.

Here we make predictions for transient rates in blind surveys (i.e. without a gravitational wave or $\gamma$-ray trigger), for orphan afterglows from short GRB jets and transients based on the expected excess as a result of low-Lorentz factor failed GRB jets \citep{2016ApJ...829..112L} and/or jets with extended wide structure \citep[e.g.][]{2017arXiv170603000L,2017arXiv170807488K,2017arXiv170807008J,2017arXiv171000275X}.
Different jet structures predict different emission properties, especially for off-axis viewing angles.
Therefore, the detection rate of orphan afterglows will give an important constraint on the structure and dynamics of the jets.
We consider only the transients from the afterglow due to the jet-ISM interaction;
such transients will be associated with all jetted short GRB progenitor models and the jet afterglow, orphan or otherwise, will have a non-thermal spectrum.
Where short GRBs are exclusively due to NS/BH-NS mergers, then additional transients will be associated;
most notably a macro/kilo-nova that will have a red/infra-red frequency peak brightness that depends on the veiwing angle, and an earlier blue/ultra-violet peak that will be apparent depending on the system inclination.
Macro/kilo-nova emission will have a thermal spectrum and a very rapid decline after the peak.

In $\S$\ref{s2} we describe the merger jet parameters and models used and in $\S$\ref{s3} we describe the method for generating the cosmological population of merger jets.
The results are described in $\S$\ref{s4}, and discussed in $\S$\ref{s5}.
Concluding remarks are made in $\S$\ref{s6}.

\section{Merger Jet Models}\label{s2}

We assume that the dominant progenitor for the short GRB population are relativistic jets from mergers \citep{2016SSRv..202...33L}.
From the observed energetics of short GRBs a luminosity function can be determined \citep[e.g.][]{2015MNRAS.448.3026W,2015ApJ...812...33S,2016A&A...594A..84G,2017arXiv171108206Z}.

We generate seven populations of merger jets where we use a \cite{2015MNRAS.448.3026W} redshift and luminosity function.
Four have homogeneous jet structure models:
\begin{enumerate}
\item[(i-iii)] WP15$_{6/16/26}$: With a coasting phase bulk Lorentz factor $\Gamma=100$ and a jet half-opening angle\footnote{Note that \cite{2016A&A...594A..84G} found this redshift distribution to indicate jet half-opening angles in the range $7^\circ\leq\theta_j\leq14^\circ$, we use the wider angle and range to include the widest observations $\theta_j\ga25^\circ$.} of $\theta_j=16\pm10^\circ$ \citep{2015ApJ...815..102F}.
For population (i) $\theta_j=6^\circ$, (ii) $\theta_j=16^\circ$, and (iii) $\theta_j=26^\circ$
\item[(iv)] LK16: With a bulk Lorentz factor distribution for the population defined as $N(\Gamma)\propto \Gamma^{-2}$ with a range $2\leq\Gamma\leq10^3$ \citep{2016ApJ...829..112L}.
We assume each jet has a half-opening angle $\theta_j=16^\circ$
\end{enumerate}
The final three jet populations use parameters that are described by the structured jet models in \cite{2017arXiv170603000L}.
These jets have a core angle $\theta_c=6^\circ$ and a wider jet component to $\theta_j=25^\circ$ in each case.
For the structured jets the luminosity function is used to determine the power within the jet core:
\begin{enumerate}
\item[(v)] LK17t, two-component jets where the wider component $\theta_c<\theta\leq\theta_j$ has energy and Lorentz factor at $5\%$ the core value
\item[(vi)] LK17p, power-law jets where the energy and Lorentz factor between the core and jet edge scale with angle from the core using a negative index power-law, $\propto (\theta/\theta_c)^{-2}$
\item[(vii)] LK17g, Gaussian jets where the energy and Lorentz factor follow a Gaussian function with angle to the jet edge, $\propto e^{-\theta^2/2\theta_c^2}$
\end{enumerate} 
The existence of a jet edge for the structured jet models is motivated by relativistic magnetohydrodynamic simulations of neutron star mergers \citep[e.g.][]{2011ApJ...732L...6R,2015PhRvD..92h4064D}

\section{Method}\label{s3}

We assume the short GRB rate\footnote{Cosmological parameters $H_0=70$, $\Omega_M=0.3$ and $\Omega_\Lambda=0.7$ are used throughout} and luminosity function given in \cite{2015MNRAS.448.3026W};
note the event rate for this distribution varies with redshift, peaking at $z=0.9$, and rapidly declines with increasing redshift ($R_{\rm GRB}=45 ~\rm{e}^{(z-0.9)/0.39}$ Gpc$^{-3}$ yr$^{-1}$ where $z\leq0.9$, and $R_{\rm GRB}=45~\rm{e}^{-(z-0.9)/0.26}$ for $z>0.9$).
At redshifts below the peak, the event rate is consistent with that found by \cite{2014MNRAS.442.2342D} and \cite{2015ApJ...812...33S}.
The luminosity function follows a broken power-law with the limits $5\times10^{49}\leq L_\gamma \leq 10^{53}$ erg s$^{-1}$ and a brake at $2\times 10^{52}$ erg s$^{-1}$;
at luminosities below the brake the power-law index is -1, and above the brake the index is -2.
The luminosity function is defined with an interval ${\rm d}\log L_\gamma$.
Note that we do not consider low-luminosity short GRBs, i.e. $L_\gamma<5\times10^{49}$ erg s$^{-1}$.
The origin of low-luminosity short GRBs is not known, the low-luminosity population could represent an extension of the usual short GRB luminosity function to lower powers or a distinct population of low-luminosity short GRBs \citep[e.g.][]{2016arXiv160603043S}.
The afterglows from a population of low-luminosity GRBs would be intrinsically very faint and the redshift distribution of the observable sample limited to `local' luminosity distances.

Using the short GRB rate (Gpc$^{-3}$ yr$^{-1}$) and luminosity function, a correlation for isotropic equivalent energy and $\nu F_\nu$ spectral peak energy $E_p$ in \cite{2013MNRAS.431.1398T}, and assuming a spectral index $\alpha=0.5$ and $\beta=2.25$ \citep{2014ApJS..211...12G} with a broken power-law, we find the minimum $\gamma$-ray luminosity for a detectable short GRB and the rate at a given redshift.
We assume a detection if the number of photons in the energy band 15-150 keV is $\geq0.3$ ph cm$^{-2}$ s$^{-1}$ \citep{2006ApJ...644..378B}.
Using the minimum observable luminosity and the short GRB rate with redshift, the all-sky number of detectable short GRBs is $\sim71$ yr$^{-1}$.
{\it Swift}/BAT detects $\sim10$ yr$^{-1}$, however as noted by \cite{2013ApJ...764..179B}, the {\it Swift}/BAT short GRB sample is contaminated by non-merger (collapsar) short duration GRBs, the fraction of merger short GRBs is $\sim60\%$. 
Using a detection rate of $\sim 6$ yr$^{-1}$ the effective field-of-view for {\it Swift}/BAT is $\sim 1.06$ sr, this is less than the BAT partially coded field-of-view $\sim1.4$ sr \citep{2013ApJS..207...19B}.
The {\it Swift}/BAT duty cycle, the sensitivity of the partially coded field-of-view, or the exact fraction of merger short GRBs may explain this discrepency.
The all-sky rate for short GRBs is used to normalize the Monte Carlo merger jet samples.

For each model (i-vii), $10^5$ merger-jets are generated.
Each jet has a random isotropically distributed inclination $i$ to the line-of-sight and a random redshift $z$ using the short GRB redshift distribution.
The jet energetics, and bulk Lorentz factor depend on the model parameters.
The prompt emission is highly beamed and only detectable for typical cosmological distances and $\gamma$-ray energies from jets inclined within the jet half-opening angle.
The $\gamma$-ray photon flux at the detector for a jet inclined within the half-opening angle is calculated considering the jet luminosity.
A correlation between the $\gamma$-ray luminosity and spectral peak energy for short GRBs is used to determine $E_p$ \citep[e.g.][]{2004ApJ...609..935Y,2009A&A...496..585G,2012ApJ...750...88Z,2013MNRAS.431.1398T}.
We use the same GRB detection criteria as that used to estimate the all-sky {\it Swift}/BAT short GRB rate.

Using the fireball model \citep{1999PhR...314..575P} and an assumed $\gamma$-ray efficiency $\eta$ for the prompt emission, the isotropic equivalent blast energy can be found from the $\gamma$-ray luminosity $L_\gamma$ and timescale $T_{90}$.
The jet kinetic energy $E_{\rm k}=L_\gamma T_{90}(1/\eta-1)$ is dissipated in shocks that form as the jet decelerates in the ambient medium giving rise to an afterglow.
The temporal evolution and peak afterglow flux of a GRB follows \cite{1998ApJ...497L..17S,1999ApJ...519L..17S}, where the peak flux is $F_p\propto n^{1/2} \varepsilon_B^{1/2} E_{\rm k} D^{-2}$, here $n$ is the ambient number density, $\varepsilon_B$ is the microphysical magnetic parameter, and $D$ is the luminosity distance.

For an off-axis observer, at an inclination greater than the jet half-opening angle $\theta_j$, the observed flux is reduced by relativistic effects.
The flux at a given observer frequency $\nu$ becomes $F_\nu(i,t) = a^3 F_{\nu/a}(0,at)$, where $a=\delta(i)/\delta(i=0)$ and $\delta=[\Gamma(1-\beta\cos i)]^{-1}$ is the relativistic Doppler factor;
$\Gamma$ is the bulk Lorentz factor, and $\beta$ is the jet velocity relative to the speed of light \citep{2002ApJ...570L..61G}.
Note that this relation is valid for a point source only and that for a jet with a defined opening angle the relativistic beaming factor for the flux is $\sim a^{2}$ for $i\la2\theta_j$, and the angle used to calculate the relativistic Doppler factor is $i-\theta_j$ where $i>\theta_j$ \citep{2000ApJ...541L..51K,2001ApJ...554L.163I}.

For all jets we use the method in \cite{2017arXiv170603000L}, with the relevant jet structure model to generate on/off-axis afterglows for the population of jets.
The ambient density is assumed to be $n=0.1$ cm$^{-3}$, microphysical parameter $\varepsilon_e=0.1$, $\varepsilon_B=0.01$, particle distribution index $p=2.5$, and $\gamma$-ray radiation efficiency $\eta=0.1$.
For each population the normalized number of {\it Swift}/BAT GRBs and orphan afterglows are counted.

Using the distribution of peak afterglows from a given model, and a transient survey telescope's per night coverage, the number of transients with or without a GRB, that have an optical counterpart brighter than the survey's detection threshold can be found.
For transients in our sample that are brighter than the LSST(ZTF) survey threshold, $r$-band magnitude $\sim24.5(20.4)$, we determine the number that are brighter than this limit for $\geq4(1)$ days.
This ensures a minimum of two detections within the proposed cadence.
For LSST we use a survey rate of $\sim3300$ deg$^2$ night$^{-1}$, covering $\sim 0.08$ of the whole sky per night;
for ZTF the survey rate is $\sim3760$ deg$^2$ hour$^{-1}$ where the average night is 8h 40m \citep{2014htu..conf...27B}.
ZTF will cover $\sim 0.09$ of the whole sky per hour, and considering the observable fraction of the sky per night, will cover $\sim22500$ deg$^2$ night$^{-1}$ or $\sim0.55$ of the whole sky per night with a 1 day cadence.

\section{Results}\label{s4}

\begin{table*}
\caption{The number of afterglow transients from a given merger jet model that are brighter than a limiting $r$-band magnitude magnitude. All models use the redshift and luminosity function from Wanderman \& Piran (2015). The GRB population in each sample is normalized to an all sky rate of {\it Swift}/BAT detectable short GRBs of $\sim 71$ yr$^{-1}$. The first value in each column is for orphan afterglows only, the values in square brackets are for GRB and orphan afterglows combined. The all-sky rates less than a given magnitude have an associated uncertainty of $\sim\pm0.7$ deg$^{-2}$ yr$^{-1}$. The LSST and ZTF detection rate is based on the mean timescale a transient is brighter than the telescope threshold}
\label{tab1}
\centering
\resizebox{\textwidth}{!}{\begin{tabular}{c c | c c c c c c c c}
 & & $\leq 26$ & $\leq 24.5$ & $\leq 21$ & $\leq20.4$ & $\langle T_{\rm{LSST}} \rangle$ & LSST &  $\langle T_{\rm{ZTF}} \rangle$ & ZTF  \\
& Model & $\times 10^{-3}$ deg$^{-2}$ yr$^{-1}$ & $\times 10^{-3}$ deg$^{-2}$ yr$^{-1}$ & $\times 10^{-3}$ deg$^{-2}$ yr$^{-1}$ & $\times 10^{-3}$ deg$^{-2}$ yr$^{-1}$ & days & yr$^{-1}$ &  days & yr$^{-1}$ \\ 
\hline
(i)& WP15$_{6}$ & 32.2 [33.6] & 25.3 [26.7] & 3.6 [5.0] & 1.4 [2.8] & 0.16 [0.20] & 13.4 [17.6] & 0.02 [0.03] & 0.6 [1.9] \\
(ii)& WP15$_{16}$ & 20.2 [22.9] & 18.5 [21.1] & 2.3 [5.0] & 0.8 [3.1] & 0.12 [0.27] & 7.3 [18.8] & 0.03 [0.06] & 0.5 [4.2]  \\
(iii)& WP15$_{26}$ & 15.7 [17.5] & 15.0 [16.7] & 3.5 [5.2] & 1.6 [3.3] & 0.11 [0.34] & 5.4 [18.7] & 0.02 [0.07] & 0.7 [5.2] \\
(iv)& LK16 & 60.0 [62.0] & 39.2 [41.2] & 3.7 [5.7] & 1.8 [3.7] & 0.54 [0.56] & 70.0 [76.1] & 0.09 [0.07] & 3.6 [5.8] \\
(v)& LK17t & 27.6 [29.3] & 21.8 [23.5] & 2.5 [4.2] & 0.9 [2.5] & 0.18 [0.30] & 12.8 [23.2] & 0.03 [0.11] & 0.6 [6.2] \\
(vi)& LK17p & 43.6 [45.4] & 29.3 [31.1] & 3.3 [5.0] & 1.3 [2.9] & 0.11 [0.25] & 10.6 [25.7] & 0.03 [0.12] & 0.9 [7.8] \\
(vii)& LK17g & 50.3 [51.5] & 34.4 [35.5] & 2.8 [4.0] & 1.3 [2.5] & 0.08 [0.14] & 9.1 [16.4] & 0.03 [0.07] & 0.9 [3.9] 
\end{tabular}}
\end{table*}

\begin{figure}
\centering
\includegraphics[width=\columnwidth]{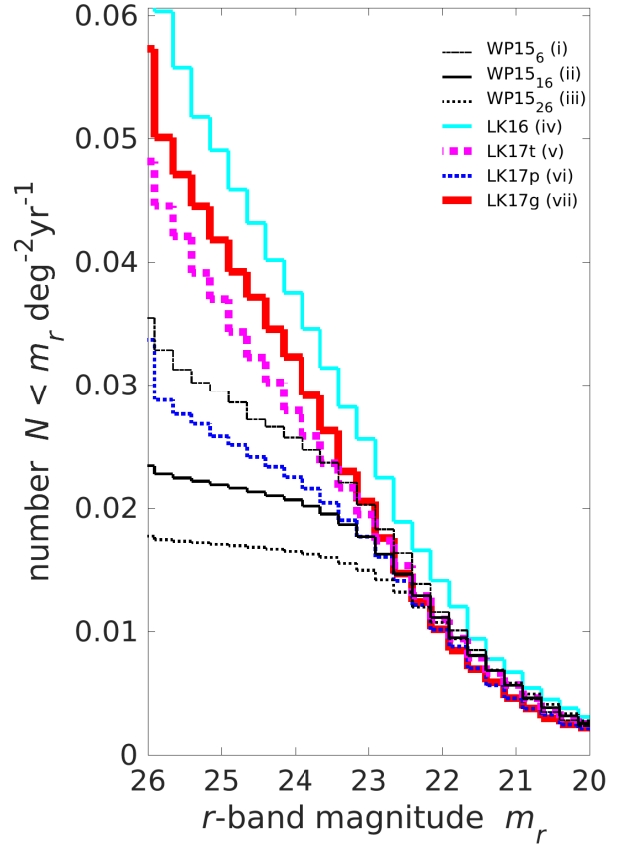}%fig1_unc.eps}%
\caption{The number of afterglows (deg$^{-2}$ yr$^{-1}$) brighter than a given magnitude for each jet model in $\Delta m_r = 0.25$ magnitude bins. The homogeneous jet models (i-iii) with a half-opening angle $\theta_j=6^\circ$ with a thin black dashed-dotted line, $\theta_j=16^\circ$ are a black solid line, and $\theta_j=26^\circ$ with a thin black dotted line. The afterglows from the low Lorentz-factor jets model (iv) are shown with a medium thickness solid cyan line. The structured jet models (v-vii) are shown with a thin blue dotted line for the two-component model, a thick pink dotted line for the power-law structured jets, and a thick red solid line for the Gaussian structured jets.}
\label{fig1}
\end{figure}

The rate of afterglow transients for each model is shown in Table \ref{tab1}.
The various models are described in \S \ref{s2}.
%Using the fraction of GRBs in each case and the derived all-sky rate of {\it Swift}/BAT short GRBs per year, the all-sky rate of mergers for each model can be found.
For LK17t, LK17p and LK17g (v-vii), the opening angle at which a GRB is detectable depends on the distance, luminosity and inclination of the source.
For the WP15 (i-iii) and LK16 (iv) models, a GRB is typically only detectable for inclinations that are less than the jet half-opening angle, $i\la\theta_j$\footnote{Merger jets at very low redshift or with very high energy may be detectable at $\gamma$-ray energies for inclinations just outside of the jet half-opening angle}.

Figure \ref{fig1} shows the number of afterglow transients per square degree per year brighter than a given $r$-band magnitude from merger jets.
Each model is indicated by a different colour and line style as described in the figure caption.
Each distribution is for both GRB and orphan afterglows (the value in square brackets in Table \ref{tab1}).
{We assume a fixed ambient density for all events of $n=0.1$ cm$^{-3}$.
However, the density for short GRB environments has a broad range and the peak flux depends on the ambient density as $\propto n^{1/2}$.
As $40-80\%$ of the short GRB population, where $\varepsilon_B=0.01$, have $n>0.1$ cm$^{-3}$ \citep{2015ApJ...815..102F}, then considering the full ambient density range, the presented results are not significantly changed.}

The homogeneous jet models with a fixed initial Lorentz factor WP15 (i,ii,iii) produce detectable GRBs, where the detector has our parameters and sensitivity, for $\sim 9\pm1\%$ of merger-jets that are inclined within the jet half opening angle.
For homogeneous jets the typical inclination for on-axis, $i<\theta_j$, orphan afterglows and/or GRB detected systems is $\sim 2\theta_j/3$;
for off-axis, $i>\theta_j$, the orphan afterglow is typically observed at an angular separation of $\sim\theta_j+1.1^\circ$, where the limiting magnitude is $\leq 26$.

%\begin{table*}
%\caption{The mean redshift.}
%\label{tab2}
%\centering
%\resizebox{\textwidth}{!}{\begin{tabular}{c c | c c c c}
% & Model & $\langle z\rangle$ & $\langle i\rangle$ GRB (deg) & $\langle i\rangle$ OA (deg) & $\langle i\rangle$ OA($>\theta_j$) (deg) \\
% (i) & WP15$_6$ & 0.47 & 3.8 & & \\
% (ii) & WP15$_{16}$ & 0.59 & 11.6 & & \\
% (iii) & WP15$_{26}$ & 0.64 & 18.8 & & \\
% (iv) & LK16 & 0.49 & 10.7 & & \\
% (v) & LK17t & 0.50 & 3.8 & & \\
% (vi) & LK17p & 0.42 & 6.0 & & \\
% (vii) & LK17g & 0.36 & 3.9 & & 
%\end{tabular}}
%\end{table*}

The typical redshift for a detected GRB with our detection criteria and parameters, for all of the jet models, is $\langle z\rangle=0.5\pm0.7$.
The uncertainty represents the different model mean values.
The measured mean redshift value for the population of {\it Swift}/BAT short GRBs with redshift is $\langle z\rangle=0.49$ \citep {2014ARA&A..52...43B}.
For the orphan afterglow populations the typical redshift is sensitive to the limiting magnitude.
Where the afterglow peaks $\leq 26$ and for the LSST sample where the peak $\leq 24.5$ then $\langle z\rangle=0.90\pm0.05$;
for the limit of $\leq 21$ and the ZTF sample with $\leq 20.4$ then $\langle z\rangle=0.80\pm0.05$ {and $\langle z\rangle=0.70\pm0.05$ respectively.} 
{For the peak magnitude limited samples, $\leq 26$ and $\leq 24.5$, the mean redshift coincides with the peak of the redshift distribution. 
For a rate that peaks at a higher redshift, the mean for the peak magnitude $\leq 26$ and $\leq 24.5$ transients is higher. 
Where a redshift distribution peaks at $z\sim1.5$, the mean is $\langle z\rangle=1.75$ and $\langle z\rangle=1.46$ respectively.}
A significant fraction of all orphan afterglows in our sample are viewed on-axis, i.e. within the jet half opening angle.
The prompt GRB can be undetected despite being favourably inclined due to the dyanmics of the jet model, the detector sensitivity, $\gamma$-ray efficiency, and/or distance.
The orphan afterglow in such a case will be phenomenolgically the same as a regular GRB afterglow \citep[e.g.][]{2013ApJ...769..130C}.

If detections are limited to two points brighter than the limiting magnitude in the given cadence of a survey telescope, then for LSST using a cadence of 4 days the number of transients brighter than magnitude 24.5 and the number for ZTF with a 1 day cadence but limiting magnitude of 20.4 is small in all cases.
For the LSST sample, with or without a GRB, the fraction of transients brighter than the threshold for the 4 day cadence considered is $\sim0.04\pm0.01$.
For ZTF this fraction is $\la0.06$ for a one day cadence.
These fractions are insensitive to the jet model.

\section{DISCUSSION}\label{s5}

We have generated a Monte Carlo distribution of merger jets for each of the jet models considered:
a population of homogeneous jets with a jet half-opening angle of $6^\circ$, $16^\circ$, or $26^\circ$;
a population of merger jets that have an independent Lorentz factor following a negative index power-law distribution;
and three structured jet models, all with a core value of $6^\circ$ and a jet edge at $25^\circ$.
The merger jets follow a \cite{2015MNRAS.448.3026W} redshift distribution for merger (non-collapsar) short GRBs and have a random isotropic inclination.
For each event the $\gamma$-ray photon flux at the detector in the energy band 15-150 keV is determined, if the flux is greater than the threshold value then a GRB is detectable.
Each population of merger-jets is normalized by the all-sky rate of {\it Swift}/BAT detectable short GRBs.

The fraction of on-axis events $i<\theta_j$ will follow the probability distribution for a randomly oriented bi-polar jet system with the jet half-opening angle $\theta_j$, i.e. $1-\cos\theta_j\sim \theta_j^2/2$.
Not all on-axis events will produce a detectable GRB or afterglows above the detection threshold.
For all models considered this is due to a combination of luminosity, distance to a merger, and spectral peak energy.
For the LK16 and structured jet models, the failed GRB fraction is higher due to suppression of the prompt emission in the low- $\Gamma$/energy jet/components.

Forward shock afterglow transients from short GRBs, on-axis failed GRBs, and off-axis orphan afterglows are detectable by both the LSST and ZTF.
The rate for both LSST and ZTF detectable transients depends on the nature of the jets in a population of mergers.
Where the jet Lorentz factor varies from jet to jet, only a small fraction of the merger jets, when viewed on-axis, will produce a detectable GRB \citep[e.g.][]{2016ApJ...829..112L}.
Afterglows are typically fainter for a population of low-$\Gamma$ failed GRBs, this is due to the later deceleration time for the jet where $t_{\rm dec} \propto \Gamma^{-8/3}$, and the lower characteristic synchrotron frequency, $\nu_m \propto \Gamma^4$, meaning the optical peak flux is lower than the maximum synchrotron flux as $\nu_{\rm obs}>\nu_m$ at the peak time.
This is reflected in the orphan afterglow rate being $\sim 2-3\times$ larger for LK16 model (iv) than for WP15$_{16}$ model (ii), where $\Gamma=100$ for all events.

For jets with structure, the orphan afterglows are typically brighter than the orphan afterglows for a population of homogeneous jets \citep{2017arXiv170603000L}.
Structured jets have higher latitude jet components with a low-$\Gamma$ that can suppress a GRB for an observer at these inclinations, thus structured jets can produce a larger fraction of orphan afterglows where the inclination is less than the jet half-opening angle.
These orphans are typically brighter than a homogeneous jet described by the $\gamma$-ray bright region of a structured jet.
However, for the two-component jet LK17t model (v) it is clear from Figure \ref{fig1} that the rate of transients is $\sim 80\%$ that of the rate for a homogeneous jet population with $\theta_j=6^\circ$, WP15$_6$ model (iii).
The two-component model will typically have a $\gamma$-ray bright core, $\theta_c=6^\circ$, and an extended `sheath' that generally fails to produce detectable GRBs.
For model (iii), $\theta_j$ is equivalent to the core size in the two-component model.
Due to the two-component models extended structure, GRBs are observable at $i>\theta_c$ in jets where the core luminosity is very high or the merger is nearby.
Therefore the fraction of GRBs from this model is larger than that for the $6^\circ$ homogeneous jet model and thus when the distributions are normalized the total number of mergers is smaller.

To consider the fraction of afterglows detected by blind sky surveys, the typical time period for which a transient is brighter than the limiting magnitude is determined.
For the LSST(ZTF) limit of 24.5(20.4) the typical timescale is shown in Table \ref{tab1}.
The product of the all-sky rate (deg$^{-2}$ yr$^{-1}$), the per night survey field-of-view (deg$^2$ day$^{-1}$), and the typical timescale for a transient (day) gives the expected rate of detectable transients for a survey.
For LSST the chance of detecting an orphan afterglow from a merger jet is reasonable, $5\la R_{\rm OA} \la 70$ yr$^{-1}$, depending on the jet model.
If we consider both orphan and GRB afterglows the rate increases, $16\la R_{\rm AG} \la76$ yr$^{-1}$.
For ZTF the rate of detected orphan afterglows from merger jets is low, $0.5\la R_{\rm OA}\la 3.6$ yr$^{-1}$.
The combined orphan and GRB afterglow rate is more promising, $2\la R_{\rm AG} \la 8$ yr$^{-1}$.
However, in each case, the afterglow transients are rarely brighter than the detection threshold for longer than the cadence. 

The differentiation between merger-jet origin orphan afterglows and collapsar or long GRB jet orphans will be difficult.
\cite{2015A&A...578A..71G} predicts a rate of $R_{\rm OA}\sim 50$ yr$^{-1}$ for the LSST from long-GRB jets.
For faint transients, the peak flux may not be brighter than the host galaxy, magnitude $\sim24-27$ \citep{2014ARA&A..52...43B}.
In such a case the {detection of a transient will depend on the survey angular resolution and the image subtraction technique}.
However, short GRBs can be hostless or have typically large offsets from the bright core or star forming regions, {but in such cases the ambient density is low and the peak flux will be fainter}.
Long GRBs are typically associated with star-forming galaxies and regions \citep[e.g.][]{1998ApJ...507L..25B,1998ApJ...508L..17D,2006Natur.441..463F}, making faint orphan transients from long GRB jets more difficult to detect.
Short GRB host galaxies systematically have an older stellar population, have a lower star-formation rate, and a higher metallicity than the host galaxies for long GRBs \citep{2014ARA&A..52...43B}.
Note however that short GRB host galaxies can be both early- and late- type galaxies.
Additionally, simulations {performed by} \cite{2017MNRAS.464.2831O} suggest that short GRB merger progenitor systems are over-produced by dwarf galaxies;
these galaxies are typically faint with surface brightness $-14\la M_{\rm B} \la -10$ \citep{2003Ap&SS.285...97S}, approximately magnitude 28-32 at $z=0.5$.
{Alternatively, by considering the natal kick velocities of NS-NS systems, the fraction of hostless short GRBs, and the dark matter potential well of galaxies, \cite{2014ApJ...792..123B} found that short GRBs are expected to be associated with galaxies that have a stellar mass $(5\pm3)\times10^{10}$ M$_\odot$.}
The differences in the host galaxy and location within the host galaxy {may} be used to distinguish between the progenitor of GRB-less transients.

{The predicted low detection rates for merger jet transients can result in confusion not only between a collapsar jet and merger jet origin but also, and perhaps dominantly, from other astrophysical transients.
Flares from active galactic nuclei (AGN), tidal disruption events (TDEs), and rapidly evolving faint supernovae (SNe) are amongst the confusion sources for fast and faint extragalactic transients.
For these events, the location within the host galaxy can help distinguish the origin, where AGN and TDEs are expected to be located within the core of a galaxy.
Spectroscopy, or colour evolution that can trace the underlying spectrum, are required to reliably distinguish between a non-thermal jet afterglow transient and a SNe or SN-like transients.
However, a thermal transient that either precedes or follows the non-thermal jet transient will indicate a merger origin, see discussion below.}

By considering the other associated transients {for a NS/BH-NS merger} i.e. resonant shattering flares or impact flares for NS mergers \citep{2013AAS...22134601T,2014MNRAS.437L...6K} or SNe for long GRBs, the origin of the orphan afterglow may be additionally constrained for nearby events.
With the development of next generation gravitational wave detectors e.g. the Einstein Telescope (ET) \citep{2010CQGra..27s4002P}, the volume within which a NS/BH-NS merger can be detected increases.
Coincident survey transients \citep[e.g.][]{2017arXiv171005845S}, within the ET detection horizon $z\sim 0.5$, and GW merger signals will be key to characterising the growing number of objects in the transient sky.
The rate of transients at $z<0.5$, with $m_r\leq26$, for our models is $(0.2\la R_{z<0.5} \la 6.8) \times10^{-3}$ deg$^{-2}$ yr$^{-1}$, and a mean rate $\langle R_{z<0.5} \rangle \sim 3.0\times 10^{-3}$ deg$^{-2}$ yr$^{-1}$.
For each model the rate is: (i) 4.5, (ii) 0.6, (iii) 0.2, (iv) 6.8, (v) 4.0, (vi) 2.4, and (vii) 5.1 where the units are $\times 10^{-3}$ deg$^{-2}$ yr$^{-1}$.
These deep survey rates for a field-of-view 3300 deg$^2$ are: (i) 3.0(22.2), (ii) 0.5(20.4), (iii) 0.2(19.6), (iv) 12.6(114.6), (v) 3.3(24.2), (vi) 2.4(44.5), (vii) 2.4(23.7) yr$^{-1}$ where the number in brackets is without the redshift condition.

The rate of orphan afterglows from long GRBs is higher than that for short GRBs due to the difference in the occurence rate of either transient.
Long GRBs typically have jet half-opening angles $\theta_j\sim 6^\circ$ \citep{2015A&A...578A..71G}.
The peak afterglow for a highly inclined system at $i>\theta_j$ decreases rapidly with increasing angle;
for long GRB jets associated with SNe, the peak flux rapidly falls below the peak of the accompanying SN\footnote{This depends on the $K$-corrected luminosity of the SN \citep{2016MNRAS.458.2973P}. GRB afterglows are brighter over a broader spectrum than SNe due to the non-thermal nature of the emission, the off-axis GRB afterglow spectrum is increasingly shifted to lower frequencies as the observation angle increases, this effectively contributes to the reduction in the observed off-axis flux for an orphan afterglow at optical frequencies} where the absolute magnitude is typically $M\sim-19$.
If the majority of long GRB jets are from core-collapse SNe then the orphan afterglow will be hidden by the SN for systems inclined at $i\ga20^\circ$ away from the jet axis \citep{2016MNRAS.461.1568K}, this will reduce the number of detectable orphan afterglows from long GRBs, where it is assumed that all long GRBs have narrow homogeneous jets.
Using the condition that an orphan afterglow from a long GRB must be inclined $i\leq20^\circ$, the fraction of the total population could be lower than the predicted $50$ yr$^{-1}$.
% is $\sim0.06$, or a minimum $\ga3$ of the predicted 50 orphan afterglows will be brighter than their accompanying SN.

\begin{figure}
\centering
\includegraphics[width=\columnwidth]{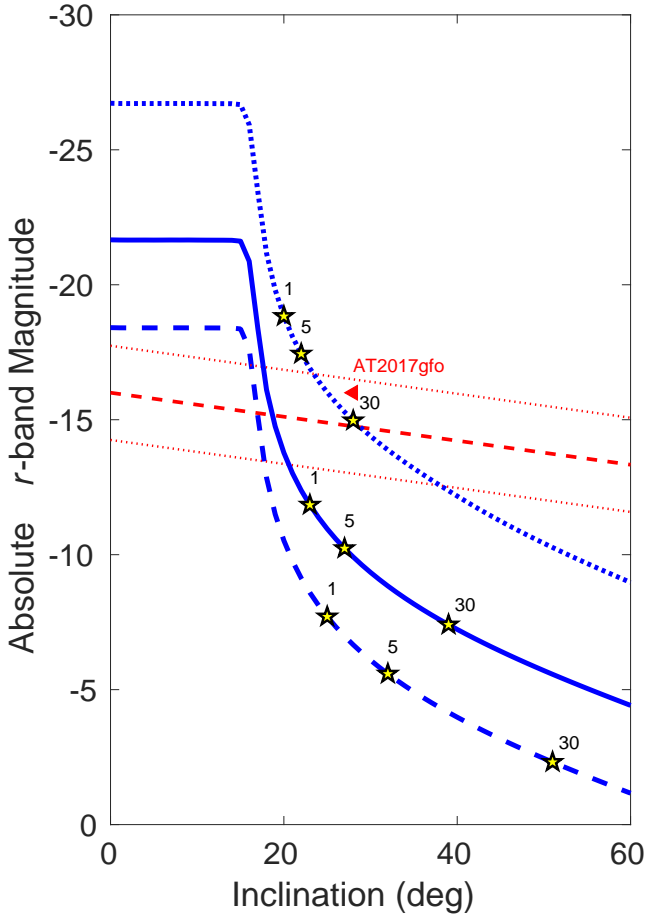}
\caption{The peak afterglow in the $r$-band with the system inclination. The blue line indictaes the peak observed $r$-band absolute magnitude at a given system inclination. The jet is a homogeneous jet with a half-opening angle of $16^\circ$ and an energy $E_{\rm iso}=5\times10^{48}$ erg for the dashed blue line $E_{\rm iso}=10^{50}$ erg for the solid blue line, and $E_{\rm iso}=10^{52}$ erg for the dotted blue line. The yellow stars indicate the peak time after the merger in days at those points. The red lines indicate the expected peak macro/kilo-nova $r$-band magnitude with inclination. The dotted red lines indicate the observed diversity of macro/kilo-nova peak fluxes.}
\label{fig2}
\end{figure}

Figure \ref{fig2} shows how the absolute magnitude for an observer in the $r$-band changes with inclination for a homogeneous jet with $\theta_j=16^\circ$ and a range of isotropic equivalent energies $5\times10^{48}\leq E_{\rm iso} \leq 10^{52}$ erg.
The peak $r$-band macro/kilo-nova flux is shown as a red line, where the range is the observed diversity \citep{2017arXiv171005442G} which agrees with the predicted range for macro/kilo-nova peak magnitudes in \cite{2013ApJ...775..113T} and \cite{2013ApJ...775...18B}.
For jets from mergers, the associated macro/kilo-nova emission, although isotropic, is generally considered to be fainter for increasing observation angles \citep[e.g.][]{2014ApJ...780...31T,2016AdAst2016E...8T,2017LRR....20....3M,2017arXiv170507084W}. 
The macro/kilo-nova decline with inclination shown in Figure \ref{fig2} assumes a linear trend from an `on-axis' view to an edge-view, where the change in magnitude is that from \cite{2017arXiv170507084W}.
The macro/kilo-nova associated with GW170817 is shown as a red triangle at an inclination of $28^\circ$ \citep[see for example:][etc.]{2017ApJ...848L..12A, 2017Sci...358.1556C, 2017arXiv171005443D, 2017arXiv171005437E, 2017arXiv171005931M, 2017ApJ...848L..34M, 2017Natur.551...67P, 2017Natur.551...75S, 2017arXiv171005850T, 2017ApJ...848L..27T}.
Macro/kilo-nova will typically peak in the $r$-band $\la 5$ days after the merger for an observer at any inclination.
However, the jet afterglow peak flux time is much later for an off-axis observer (days to months) than for an on-axis observer (minutes to hours).
Where an orphan afterglow peaks $\la 1$ day or $\ga 5$ days, a survey telescope may have two opportunities to observe the merger, one from the afterglow and a second from the macro/kilo-nova;
the afterglow transient will typically fade much slower than that from a macro/kilo-nova and will have a non-thermal spectrum.
Thus, even if the transients have coincident peaks, the afterglow will fade more slowly than the macro/kilo-nova.
Additionally, due to the broadband nature of the afterglow a radio transient should accompany the optical, possibly peaking at a later time depending on the inclination of the system.

The orphan afterglow population may be dominated by low-luminosity GRBs.
Such low-luminosity GRBs form a distinct population \citep{2010MNRAS.406.1944W} where the rate is greater than that for long GRBs.
The afterglow from low-luminosity GRBs is fainter than that for long GRBs \citep{2015MNRAS.448..417B}, but the lower Lorentz factor of the ejecta means that any off-axis emission will have a reduced beaming effect.
Where an orphan afterglow is brighter than the accompanying SNe, then low-luminosity GRB orphan afterglows may dominate the blind survey population.

The jets that produce long GRBs may exhibit the same structure or dynamical diversity as proposed for merger jets, the number of detectable orphan afterglows from long GRB or collapsar jets will be higher than that predicted by assuming homogeneous structured jets.
The increased rate of orphan transients from either collapsar jets or merger jets would indicate the presence of intrinsic jet structure or a dominant population of low-$\Gamma$ jets.
If long GRB jets follow the latter, i.e. a dominant low-$\Gamma$ population \citep{2014ApJ...782....5H}, the rate of orphan afterglows from collapsar jets would be higher by a similar fraction to that demonstrated here for merger jets with the LK16 model (iv).
The lightcurve of an on-axis orphan afterglow will appear phenomenologically the same as a GRB afterglow, i.e. a power-law decay, $\propto t^{-1}$ with an observable break at late times.
Whereas an off-axis orphan afterglow would decay with a steep, $\propto t^{<-2}$, decline with no jet-break.

We used a population of mergers that follow a lognormal time delay redshift distribution \citep{2015MNRAS.448.3026W}.
If NS/BH-NS mergers follow a power-law time delay distribution that peaks $z\sim1.5-2$, then a higher fraction of the short GRB population would go undetected due to the large luminosity distance.
The observation of a host for short GRB 111117A at $z=2.211$ could challenge the lognormal time delay model, although this redshift is still within the lognormal limits, the probability is $\sim2$ orders of magnitude lower than for the peak at $z=0.9$ \citep{2017arXiv170701452S}.
Detectable orphan afterglows from a power-law time delay redshift distribution will follow the rates predicted here {where $m_r\lesssim23$}.
A significant excess would exist at very faint magnitudes, $m_r>26-28$, where the population distribution peaks at a redshift $z\ga1.5$.
{The peak of the redshift distribution can be traced by the faint transients.}
For a discussion of a short GRB population with such a distribution see \cite{2015ApJ...812...33S} and \cite{2016A&A...594A..84G}.

\section{CONCLUSIONS}\label{s6}

We have shown that the rate of orphan afterglows from merger (non-collapsar) short GRBs detectable by the LSST is $5\la R_{\rm OA} \la 70$ yr$^{-1}$, where the rate is $\sim 7.3$ yr$^{-1}$ for a population of homogeneous jets with $\theta_j=16^\circ$.
Where GRB afterglows are included, the rates become $16\la R_{\rm AG}\la76$ and $\sim19$ yr$^{-1}$ for homogeneous jets with $\theta_j=16^\circ$.
The ZTF detection rate for orphan afterglows from short GRBs is low $0.5\la R_{\rm OA} \la 3.6$, the rate for afterglows with or without a {\it Swift}/BAT detectable GRB is $2\la R_{\rm AG} \la 8$ yr$^{-1}$, and $\sim 4.2$ yr$^{-1}$ for a population of homogeneous jets with $\theta_j=16^\circ$.

For populations of jets narrower than $\theta_j=16^\circ$, the rate of orphan afterglows increases.
For LSST the orphan afterglow rate from a population of narrow short GRB jets is $\sim 13.4$ yr$^{-1}$.
This increase is due to the increased rate in the parent merger population due to the normalization required to produce the detectable all-sky {\it Swift}/BAT non-collapsar GRB rate.

If the population of jets that produce short GRBs is dominated by jets with a low-$\Gamma$, then the rate of orphan afterglows will increase significantly, $\sim 70$ yr$^{-1}$ for LSST and $\sim 3.6$ yr$^{-1}$ for ZTF.
Where these jets result in failed-GRBs and are viewed within the jet half-opening angle, the lightcurve will appear phenomelogically the same as an on-axis GRB afterglow lightcurve.

If jets exhibit intrinsic structure, where the jet energetics extend beyond a homogeneous core to a defined edge, then the rate of orphan afterglows is greater than that for a homogeneous population;
with the exception of the two-component jet structure when compared to the narrowest homogeneous jet population.

If LSST modifies the observation strategy from a fast survey to a deep-drilling field, then the obtainable sensitivity will increase.
By focusing on a single field the potential to detect the same transient with multiple observations increases and with this the ability to identify orphan afterglows.
As shown in Figure \ref{fig1} and Table \ref{tab1}, the rate of detectable merger-jet transients increases significantly from a limiting magnitude of 24.5 to 26 for a population of narrow jets, a jet population dominated by low-Lorentz factors, or jets with intrinsic structure.

The structured or dynamical models tested here could equally be applied to collapsar or long GRB jets.
The observed rate of orphan afterglows from such jets would increase by a similar fraction for each case.
Careful filtering of transients is required to successfully identify an orphan afterglow from either short or long GRB jets.
Orphan afterglows fade rapidly and will rarely be above the detection threshold $>1$ day, single point candidate identification and fast targeted follow-up will be required. 

\section*{Acknowledgements}
This research is supported by STFC grants.
The authors thank the anonymous referee for helpful comments, and Samaya Nissanke and the participants of the Nordita `The Physics of Extreme Gravity Stars' workshop and conference 2017 for useful discussions.
GPL thanks Simon Prentice for helpful comments.

\bsp
\label{lastpage}
\end{document}